# Dispersion mapping as a simple postprocessing step for Fourier domain Optical Coherence Tomography data

Sylwia M. Kolenderska[1*], Bastian Bräuer[1], and Frédérique Vanholsbeeck[1]

[1]The Dodd-Walls Centre for Photonic and Quantum Technologies, Department of Physics, Auckland 1142, New Zealand
[*]skol745@aucklanduni.ac.nz

## ABSTRACT

Optical Coherence Tomography (OCT) was originally conceived as a volumetric imaging method. Quickly, OCT images went beyond structural data and started to provide functional information about an object enabling for example visualization of blood flow or tissue elasticity. Minimal or no need for system alterations make functional OCT techniques useful in performing multimodal imaging, where differently contrasted images are produced in a single examination. We propose a method that further extends the current capabilities of OCT and requires no modifications to the system. Our algorithm provides information about the sample's Group Velocity Dispersion (GVD) and can be easily applied to any OCT dataset acquired with a Fourier domain system. GVD is calculated from the difference in material's optical thickness measured from two images obtained for different spectral ranges. Instead of using two separate light sources, we propose to apply a filter-based, numerical procedure that synthesizes two spectra from one broadband spectrum. We discuss the limitations of the method and present GVD values for BK7 and sapphire and ocular media: cornea and aqueous humour of a rat eye. Results corroborate previous measurements using two different light sources.

## Introduction

Optical Coherence Tomography (OCT) is a non-invasive and non-contact method for volumetric imaging with micrometric axial and lateral resolution[1–3]. It uses a Michelson's interferometer to generate a depth profile of a sample at one lateral point, an A-scan, by producing interference whenever the optical path length in the reference arm equals the optical path set by a layer in the sample in the object arm. In Time domain OCT (Td OCT), an A-scan is directly obtained by axially translating the reference mirror and detecting the signal with a photodiode. A better solution in terms of speed and sensitivity is Fourier domain OCT (Fd OCT), where the reference mirror's position is fixed and corresponds to the position of the sample in the object arm. In Fd OCT, an A-scan is produced by Fourier transformation of a spectrum detected by a spectrometer in Spectral OCT (SOCT) or by a photodiode while the source's spectrum is swept in time in Swept Source OCT (SS OCT).

OCT can provide much more information about the sample than structural characteristics[4]. Many functional extensions to OCT now exist: Doppler OCT[5–7] for visualisation of blood flow in eye, kidneys, skin etc., Polarization-sensitive OCT (PS OCT)[8–14], where tissue changes due to burns or neoplasms can be observed because of the birefringence they induce, or Optical Coherence Elastography (OCE)[15–20] for visualizing the change in elasticity of a sample. Functional OCT techniques use different contrast mechanisms, birefringence in PS OCT or elastic modulus in OCE, to identify sections of a sample with different properties and require little alteration to the system or, in the vast majority of cases, no alterations at all. Techniques that don't add any complexity to the system are very much sought for as they can be universally used for multimodal imaging and, thus, provide complementary data in a single examination[21–25].

Group Velocity Dispersion (GVD) has also been suggested as a method for material differentiation in OCT[26], but in the very first application, sample thicknesses of several millimetres were required; thicknesses which exceeded the typical imaging range of an OCT instrument (1-2 mm). However, recently, new approaches for chromatic dispersion measurement have been demonstrated and can be broadly categorized as time of flight or point spread function (PSF) broadening measurements. For example, on one hand, GVD can be determined by estimation of depth-dependent resolution degradation with an analysis of image speckle[27], which is a very good solution for characterization of highly scattering media. On the other hand, in a method presented in Ref.[28], two light sources are used to generate two OCT images, which are then analysed to extract the optical thicknesses of imaged media. By comparing these optical thicknesses, GVD can be calculated. This method proved to be successful at determining GVD values in layers as thin as 160 $\mu$m and, most recently, allowed dispersion measurements in ocular media such as cornea and aqueous humour in rat eyes[29].

The experimental configurations in Ref[28] and Ref[29] included two separate, synchronized, swept laser sources at different wavelengths. In these experimental configurations, it is necessary to align light beams coming from the two sources so that they propagate co-linearly towards a sample. What's more, the fringes are detected by two different photodiodes. These characteristics of the experimental configuration may result in the two images not being perfectly superimposed, and subsequently cause inaccuracies in the measurement.

Here we present a purely post-processing method for GVD-calculation-based material differentiation. This method eliminates alignment issues and can be applied to data from any Fourier domain OCT system. We use different glasses of different thicknesses to validate this new approach, and to assess its limitations. Finally, the results for the cornea and aqueous humour of a rat eye are presented.

## Materials and methods

The methods discussed in Ref.[28] and Ref.[29] and here are conceptually similar to the approach presented in Ref[30]. The authors compare optical thicknesses of layers of water, as well as cornea, aqueous humour and lens in a rat eye, for two different wavelengths. However, in the method of Ref.[30], the difference in optical thickness in conjunction with an estimate of a geometrical thickness is used to determine the change of group refractive index, which then can be easily recalculated to GVD.

In our new approach to the method, two spectra are synthesized by applying two Gaussian filters to the original spectrum. The two filters are centred around two different wavelengths, $\lambda_1$ and $\lambda_2$ (corresponding to two frequencies, $\omega_b$ and $\omega_a$) in the original spectrum and their total bandwidth doesn't exceed the total spectral bandwidth of the original spectrum (Fig. 1a). The resulting sub-spectra are then independently Fourier transformed to produce two depth profiles of the same place in a layered sample. The optical thickness of a given layer in the sample will be different for these two A-scans, because they correspond to two different spectral regions (Fig. 1b). The walk-off, $\Delta z_{ab}$, is determined from the difference in the optical thicknesses and subsequently used to calculate GVD:

$$\beta_2 = \frac{1}{c\Delta\omega_{ab}}\frac{\Delta z_{ab}}{l_s} \quad (1)$$

where $l_s$ is the geometrical thickness of the material, $\Delta\omega_{ab} = \omega_b - \omega_a$ and corresponds to a wavelength distance $\Delta\lambda_{ab} = \lambda_2 - \lambda_1$, c – the speed of light in vacuum, and $\beta_2$ is a GVD calculated at the midpoint between the two frequencies. Because the midpoint is defined in frequencies, its wavelength counterpart is not an arithmetic average of $\lambda_1$ and $\lambda_2$, but can be easily calculated with the following formula: $2\lambda_1\lambda_2/(\lambda_1 + \lambda_2)$.

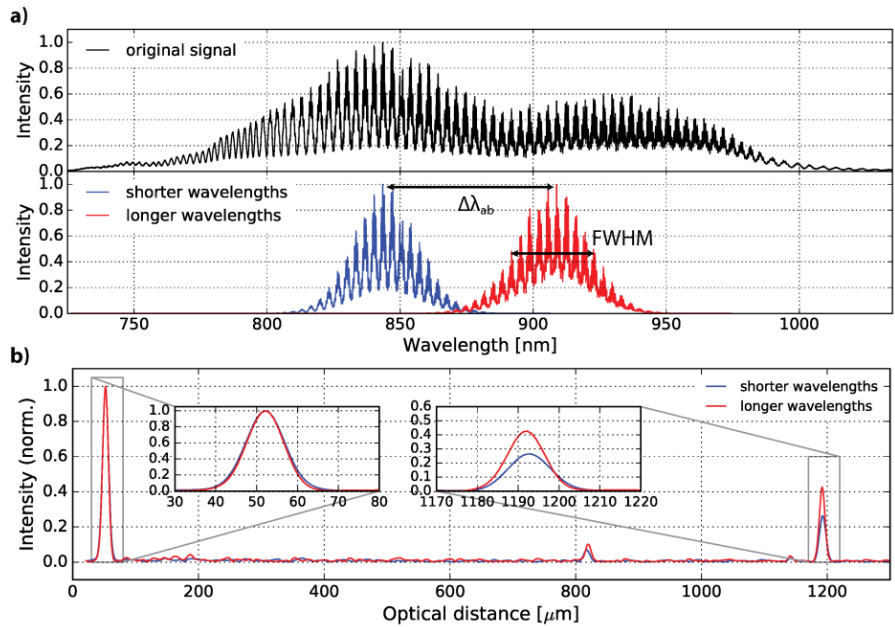

**Figure 1. Method for extraction of $\beta_2$ value from a spectrum.** The original spectrum (**a**, top) is multiplied by two Gaussian filters so that two spectra in two different spectral regions are obtained (**a**, bottom). (**b**) A-scans produced by Fourier

transformation of these two spectra show different optical thicknesses of the same layer, what enables calculation of the walk-off, $\Delta z_{ab}$, and subsequently, $\beta_2$ itself.

Since OCT provides information about optical distances, the sample thickness, $l_s$, is unknown. For glass objects used in our study, the precise geometrical thicknesses are given by the manufacturer. However, for the ocular media, the thicknesses have to be estimated. In order to do so, first, the optical thickness is determined from the A-scan obtained by Fourier transformation of the original spectrum. This optical thickness is then divided by an approximated value of group refractive index: 1.380 for cornea (estimated from Ref.[30]) and 1.341 for aqueous humour (group refractive index of water[31]).

To validate the method, we use a Spectral OCT system. The light from a supercontinuum light source, Leukos Pegasus, is filtered to have a central wavelength of 875 nm and full width at half maximum (FWHM) of 130 nm. It passes through a Michelson-Linnik interferometer with achromatic lenses with focal length, f, of 40, 50 and 60 mm in each arm. The signal is then coupled by a single-mode fibre (AFW Technologies FOP830-LB-5-SM800-22), which transmits it to a home-built spectrometer. In the spectrometer, the interfered signal is collimated by a lens (f=75 mm), dispersed by a grating (1200 lines/mm) and then the resulting spectrum is imaged by the second lens (f=200 mm) onto the camera (Basler Sprint spL8192-70km, 8192 pixels, 10$\mu m$ pitch). Detected spectra are resampled from wavelengths to wavenumbers and undergo numerical dispersion compensation before they are Fourier transformed[32].

## Limitations of the method

To check the parameters of the system, we used a mirror as an object. The FWHM of the PSF obtained with the original spectrum is 3.5 $\mu$m. We filter the original spectrum so that the shorter-wavelength sub-spectrum has a central wavelength of 840 nm and its FWHM is equal to 30 nm, and the longer-wavelength sub-spectrum is centred around 900 nm and its FWHM is also 30 nm. The midpoint is 868.97 nm (see the text under Eq. 1) and the wavelength distance between the peaks, $\Delta\lambda_{ab}$ – 60 nm (Fig. 1a). The measured PSF's FWHM for the sub-spectra is 13.2 $\mu$m and 14.7 $\mu$m, respectively. The two PSFs are positioned at the same place in the A-scans for the first interface (Fig. 1b), because the original spectrum is interpolated to convert wavelengths to wavenumbers before filtering and numerically dispersion-compensated before filtering.

The choice of the central wavelengths and the bandwidths that are filtered out from the original spectrum is dictated by several criteria. First, the orders of magnitude of values of $\beta_2$ and the thickness of the layer under investigation have to be considered. Too small value of either of these parameters could produce a walk-off that may not be detectable if the spectral separation of the filtered spectra is not big enough (see Eq. 1). The minimum walk-off that can be measured with an OCT is related to the resolution of the process of Fourier transformation (FT). The resolution of FT is the minimum difference in frequency which can be distinguished with FT, which for OCT is the minimal distinguishable distance between two peaks in the A-scan. Simple calculations based on basic properties of FT show that the minimum walk-off, $\Delta z_{min}$, depends on the refractive index of the medium, $n$, and the total bandwidth of the spectrum, $\Delta\omega_{tot}$, that is Fourier transformed:

$$\Delta z_{min} = \frac{c}{2n\Delta\omega_{tot}} \qquad (2)$$

$\Delta\omega_{tot}$ is the total bandwidth of a sub-spectrum. However in practice, the total spectral bandwidth of the original spectrum should be used as $\Delta\omega_{tot}$ in calculations, because the sub-spectra are filtered out from the original spectrum by multiplying it by a window of a Gaussian form, which is non-zero over the entire spectrum and therefore doesn't limit the sub-spectra's total bandwidth.

This information can be used to determine a criterion for the selection of parameters $\Delta\omega_{ab}$ and $\Delta\omega_{tot}$ for given parameters of a medium: $n$, $\beta_2$ and $l_s$:

$$\Delta\omega_{ab}\Delta\omega_{tot} \geq \frac{1}{2n\beta_2 l_s} \qquad (3)$$

The accuracy of the technique is further limited by the noise in the system that induces slight fluctuations of the position of the peak in the A-scans over time. These fluctuations produce different optical thickness values and, as a result, slightly different values of $\beta_2$. These values have a Gaussian distribution whose FWHM depends on two parameters that can be controlled in the method: total spectral bandwidth of the sub-spectra and the number of FT points.

To see the influence of these two parameters on the results, we acquired 300 spectra at one point of a 425$\mu m$ thick sapphire glass slide. $\beta_2$ was calculated for every measured spectrum for FWHM varying from 15 nm to 40 nm for each sub-spectra with $\Delta\lambda_{ab}$ of 60 nm and a total number of FT points equal to 524,288. The sample histograms of $\beta_2$ values for FWHM of 20 nm, 30 nm and 40 nm are presented in Figure 2a, 2b and 2c, respectively. The values of standard deviation of the obtained distributions were plotted as a function of the total bandwidth of each sub-spectra in Fig. 2d.

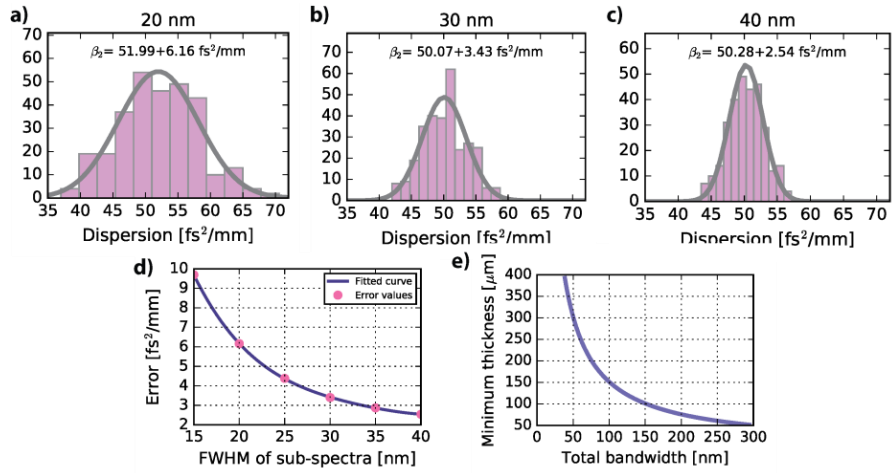

**Figure 2. Limitation coming from the total bandwidth of the sub-spectra.** $\beta_2$ values obtained from 300 A-scans taken at the same place of a 425 μm thick sapphire glass for two sub-spectra having a total bandwidth of (**a**) 20 nm, (**b**) 30 nm and (**c**) 40 nm. (**d**) The experimental data showing the decrease of the error of determination of $\beta_2$ with the increase of the bandwidth of the sub-spectra. (**e**) Minimum thickness of the sapphire glass for which $\beta_2$ can be calculated with this method as a function of the total bandwidth of sub-spectra. $\Delta\lambda_{ab}$ and the number of FT points were constant throughout the calculations, 60 nm and 542,288 (=$2^{19}$), respectively. The reference value of GVD for sapphire is 49.9 fs$^2$/mm[33].

The error clearly decreases with the bandwidth. The increase in bandwidth improves the axial resolution, which means that the true location of a boundary in an object can be determined more accurately in the A-scan. The situation is equivalent to the energy-time uncertainty principle: the wider the distribution of the energies, i.e. the wider the spectrum of light, the narrower the distribution of time, i.e. the location in the A-scan and, consequently, the determined thicknesses, the walk-off and, eventually, the dispersion.

The minimum detectable walk-off (Eq. 2) determines the minimum thickness of a material for which GVD can be calculated. It decreases as a function of the total spectral bandwidth is presented in Figure 2e.

Finally, the sub-spectra have to be padded with a sufficient number of zeros to ensure accurate determination of the peaks' maxima after FT. The histograms presenting the distribution of dispersion values for a 425μm thick sapphire glass for spectra zero-padded to contain 131,072 ($2^{17}$), 524,288 ($2^{19}$) and 2,097,152 ($2^{21}$) points are presented in Fig. 3a, 3b and 3c, respectively.

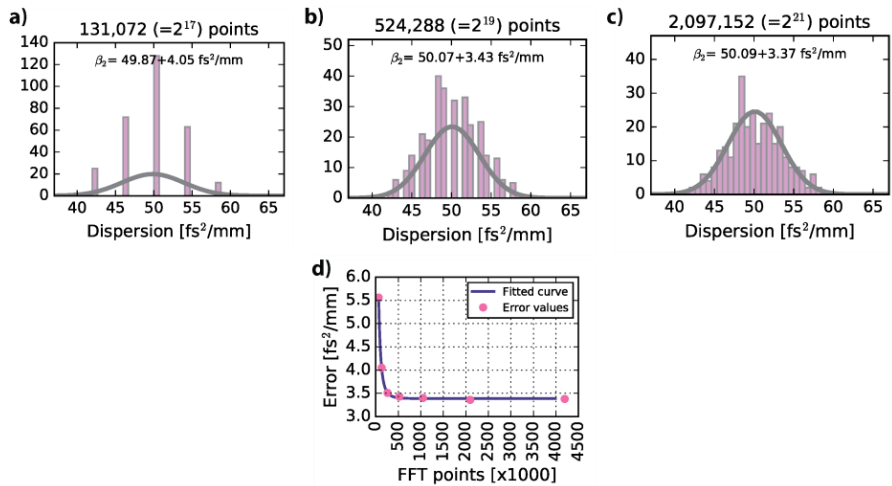

**Figure 3. Limitation coming from number of FT points.** Distributions of $\beta_2$ values obtained from 300 A-scans taken at the same place on a 425 $\mu$m thick sapphire glass for two sub-spectra being zero-padded to have (**a**) 131,072 (**b**) 524,288 and (**c**) 2,097,152 points for Fourier transformation. (**d**) Dependence of the error of $\beta_2$ determination on the number of points during Fourier transformation. Insufficient number of points can cause the increase of the error due to insufficient accuracy in determining the optical thickness in A-scans. $\Delta\lambda_{ab}$ and $\Delta\lambda_{tot}$ were constant throughout the calculations and are equal to 60 nm and 30 nm, respectively. The reference value of GVD for sapphire is 49.9 fs$^2$/mm[33].

In the plot in Fig. 3d depicting the error as a function of FT points, we can see that the error increases substantially if the zero-padding is not optimal. Because the number of bins in the histograms in Fig. 3 is fixed, a discreet distribution of $\beta_2$ values can be seen. It is due to the fact that the accuracy in determining the walk-off is one pixel, which corresponds to 0.0788 $\mu$m, 0.0197 $\mu$m and 0.0049 $\mu$m for zero-padding of 131,072, 524,288 and 2,097,152 points, respectively, and, as a result, to the accumulation of 4.19 fs$^2$/mm, 1.05 fs$^2$/mm and 0.26 fs$^2$/mm in the $\beta_2$ value, whereas the bin size in the histograms is 0.67 fs$^2$/mm.

## Results

We chose to perform a Fourier transformation on 262,144 (2$^{18}$) points to ensure a sufficiently accurate determination of the thickness of the layers, which in our case is 0.0364 $\mu$m per pixel, for sub-spectra centred around 840 and 900 nm (thus separated by $\Delta\lambda_{ab}$=60 nm) with an FWHM of 30 nm each. The mid-point between the peaks, for which the GVD value is determined, is 868.97 nm (see the text under Eq. 1).

### Glass
We started with measuring dispersion values for two double-layered objects:
- 425 $\mu$m thick sapphire glass on top of 975 $\mu$m thick BK7 glass,
- 140 $\mu$m thick BK7 glass on top of a 975 $\mu$m thick BK7 glass.

20 A-scans over a lateral distance of 200 $\mu$m on the sample were acquired and analysed in terms of $\beta_2$. Then the dispersion value was calculated as a mean over the dispersion values for each acquired A-scan. The results are presented in Table 1.

| Object # | Material | Thickness [$\mu$m] | Dispersion [fs$^2$/mm] |
|---|---|---|---|
| 1 | sapphire | 425 | 50.5 ± 4.22 |
| 1 | BK7 | 975 | 38.7 ± 3.3 |
| 2 | BK7 | 140 | 38.8 ± 18.2 |
| 2 | BK7 | 975 | 37.1 ± 2.9 |

**Table 1. Dispersion values for sapphire and BK7 at $\lambda$=868.97 nm.** $\beta_2$ is calculated as a mean of values obtained from 20 A-scans covering lateral distance of 200 $\mu$m on the sample. The reference values: 38.3 fs$^2$/mm for BK7[33] and 49.8 fs$^2$/mm for sapphire[33].

The standard deviation of $\beta_2$ is larger for thinner samples, because the error of determination of a walk-off is approximately constant – at least when induced only by random fluctuations – and therefore makes the ratio $\Delta z_{ab}/l_s$ in Equation 1 larger for thinner materials. For 262,144 Fourier transform points and assuming these random fluctuations as the only source of error, the standard deviation of the walk-off is around 2 pixels, which is 0.0788 $\mu$m and corresponds to 1.80 fs$^2$/mm for a 975 $\mu$m thick material, 4.14 fs$^2$/mm for 425 $\mu$m and 12.57 fs$^2$/mm for 140 $\mu$m.

### Rat eye
Next, we went to investigate the dispersion properties of a rat eye. Figure 4a and 4b show sample images obtained after Fourier transforming the shorter and longer wavelength regions of the original spectra. The green rectangle indicates 180 $\mu$m by 2.11 mm area of the eye containing 12 A-scans which were used for the dispersion calculation. The most central areas were chosen, so that the error related to refraction on curved surfaces is minimized. This error is due to the fact that two light beams of different wavelengths passing obliquely through a boundary are refracted at different angles and therefore traverse different distances in the sample acquiring a larger or smaller walk-off.

The geometrical thickness of the cornea and the aqueous humour calculated as a mean over all 12 A-scans is 131.1 ± 4.7 µm and 373.6 ± 4.0 µm, respectively. The expected standard deviation of dispersion value for these layers – assuming only random fluctuations in the system as a source of error – is 13.42 fs$^2$/mm for the cornea and 4.72 fs$^2$/mm for the aqueous humour, which is why we decided to detect maximum values of peaks that correspond to the boundaries of the two layers (blue dots in Fig. 4c and red dots in Fig. 4d and a sample A-scan in Fig. 4e) in order to determine optical thickness rather than use a segmentation algorithm and then fit a polynomial as this process may introduce additional error. $\beta_2$ values for cornea and aqueous humour are presented in Table 2.

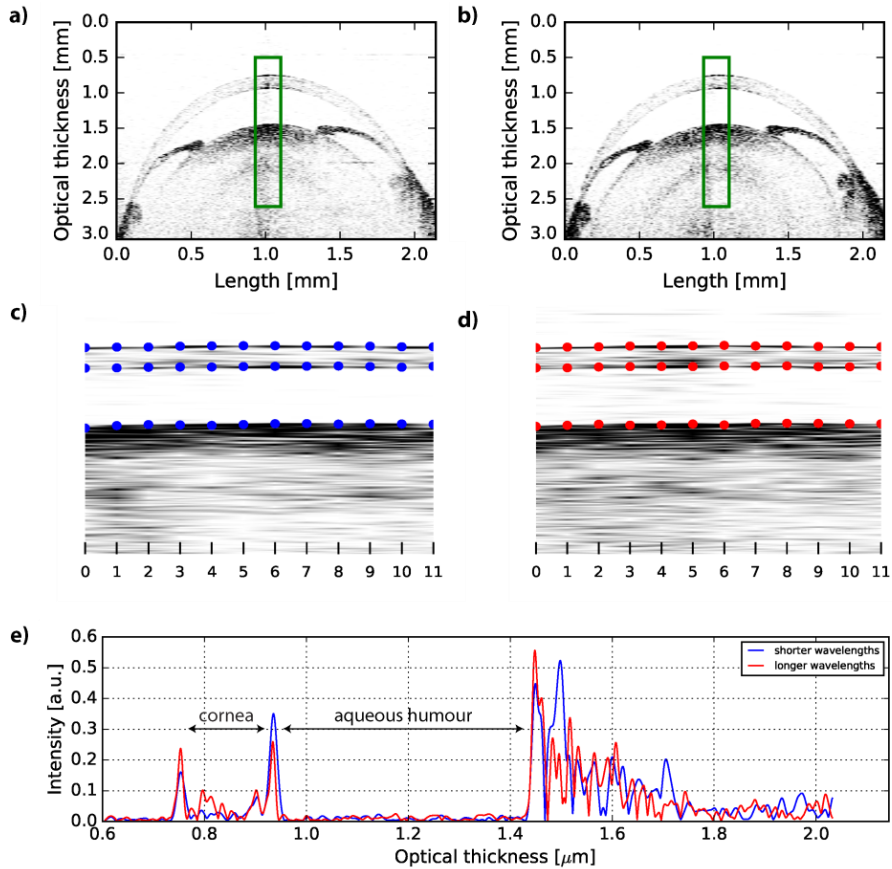

**Figure 4. Extraction of $\beta_2$ from the images of a front of a rat eye.** An image of a rat eye corresponding to (**a**) the shorter wavelength range, (**b**) the longer wavelength range in the original spectrum. Extracted 180 µm x 2.11 mm region for (**c**) shorter and (**d**) longer wavelengths with maximum values for the peaks representing the boundaries of cornea and aqueous humour (blue dots in (**c**) and red dots in (**d**)). (**e**) The 7th A-scans from (**c**) and (**d**) show different optical thicknesses of cornea and aqueous humour in the eye.

|  | Thickness [µm] | Dispersion [fs$^2$/mm] |
|---|---|---|
| Cornea | 131.1 ± 4.7 | 122.7 ± 19.7 |
| aqueous humour | 373.6 ± 4.0 | 23.5 ± 5.3 |

**Table 2. Dispersion values and thickness for cornea and aqueous humour of a rat eye for 868.97 nm.** $\beta_2$ is calculated as a mean of values obtained from 12 A-scans covering lateral distance of 180 µm on the eye.

## Discussion

We have shown that with a simple numerical procedure of filtering the spectra acquired in Fourier domain OCT and comparing optical thicknesses in the resultant A-scans a $\beta_2$ value of an imaged object can be determined. Calibration performed on materials with well-defined properties revealed that several parameters have to be optimised for the method to be successful. Firstly, the total spectral bandwidth of the sub-spectra influences the value of minimum measurable walk-off and the size of standard deviation of the results. Next, the number of points used for FT, if too small, can decrease the accuracy of determining optical thicknesses and substantially increase the error. Finally, the lower limit is set by noise in the system which causes slight fluctuations of the measured optical thicknesses over time and has the most negative impact for thinner materials. The method enables determination of dispersion values for ocular media: cornea and aqueous humour. The mean value calculated for cornea, 122.7 fs$^2$/mm, is approximately 6 times bigger than for water and a little bigger than the value reported in Ref.[30] for the similar spectral range, 108.5 fs$^2$/mm. The mean dispersion value calculated for aqueous humour, 23.5 fs$^2$/mm, is similar to that of water and slightly bigger than the value in Ref.[30] for the similar spectral range: 16.8 fs$^2$/mm. The same magnitudes of the dispersion value ratios for the cornea and the aqueous humour compared to water were reported in Ref.[29] at 1200 nm.

**Acknowledgements**

This work was supported by Marsden fund contract number UoA1509. We would like to thank Yadi Chen for providing the eye samples, Dr. Ehsan Vaghefi for all the interesting discussions and Dr Cushla McGoverin and Dr Stuart Murdoch for careful proofreading.

## Author contributions statement

B.B. conceived the idea of using one source for determination of dispersion. S.K. developed the idea further by suggesting using numerical methods and carried out the experiments. F.V. supervised the work and obtained funding for the work. S.K. and F.V. wrote the manuscript and refined the theory. All authors analysed the results and reviewed the manuscript.

## Additional information

Competing financial interests: The authors declare no competing financial interests.